\documentclass[twocolumn,preprintnumbers,amsmath,amssymb,superscriptaddress]{revtex4}

\usepackage{graphicx}
\usepackage{dcolumn}
\usepackage{bm}
\usepackage[usenames]{color}

\begin{document}


\title{Substrate gating of contact resistance in graphene transistors}

\author{Dionisis Berdebes}
\email{dionisis@purdue.edu}
\affiliation{School of Electrical \& Computer Engineering, Purdue University, West Lafayette, IN47906, USA }%

\author{Tony Low}
\affiliation{School of Electrical \& Computer Engineering, Purdue University, West Lafayette, IN47906, USA }%

\author{Yang Sui}
\affiliation{School of Electrical \& Computer Engineering, Purdue University, West Lafayette, IN47906, USA }%

\author{Joerg Appenzeller}
\affiliation{School of Electrical \& Computer Engineering, Purdue University, West Lafayette, IN47906, USA }%

\author{Mark Lundstrom}
\affiliation{School of Electrical \& Computer Engineering, Purdue University, West Lafayette, IN47906, USA }%



\date{\today}

\begin{abstract}

Metal contacts have been identified to be a key technological bottleneck for the realization of
viable graphene electronics. Recently, it was observed that for structures that possess both
a top and a bottom gate, the electron-hole conductance asymmetry can be modulated by
the bottom gate. In this letter, we explain this observation by postulating the presence of 
an effective thin interfacial dielectric layer between the metal contact and the underlying  graphene.
Electrical results from quantum transport calculations accounting for this 
modified electrostatics corroborate well with the experimentally measured contact resistances. Our study 
indicates that the engineering of metal-graphene interface is a crucial step 
towards reducing the contact resistance for high performance graphene
transistors.

\end{abstract}

\maketitle
%
Since its experimental isolation \cite{novoselov01,novoselov02,zhang01}, graphene has attracted significant attention from the scientific community, due to its unique electronic structure and physical properties \cite{geim01,neto09}. Its excellent transport properties and the ability to tune the carrier concentration with electrical gates, also makes it a material with great technological promise. Potential applications range from RF devices and transistors \cite{avouris01,lin10,schwierz01} to bio-sensors \cite{schedin07} and flexible electronics \cite{kim08}. The metal-graphene contact, is however, a key technological challenge for graphene-based electronic devices.  For current-generation silicon metal-oxide-semiconductor field-effect transistors (MOSFETs), the International Technology Roadmap for Semiconductors calls for a resistance of $80 \,\Omega$-$\mu  m$ per contact, which is about 10\% of the transistor's on-resistance $V_{DD}/I_{ON}$ \cite{roadmap01}.  Graphene's excellent transport properties should produce transistor on-resistances considerably lower than those of silicon MOSFETs.  To realize the performance potential afforded by the excellent transport properties of graphene, exceptionally low contact resistances will be required \cite{schwierz01,avouris01}.  It is, therefore, essential to develop a thorough understanding of metal-graphene contacts and of the fundamental lower limits for the contact resistance.  In this paper, we develop a model that explains the recently observed substrate modulation of contact resistance in graphene transistors \cite{chen01,kim02}.  We argue that this effect is due to the presence of an effective thin metal-graphene interfacial dielectric layer. Using this model, we estimate two important components of the series resistance and establish lower bounds for the contact resistance.  The study provides an improved understanding of the metal-graphene contact that may prove useful for improving device performance.

\begin{figure}[h!]
\centering
\scalebox{0.55}[0.55]{\includegraphics*[viewport=180 80 630 550]{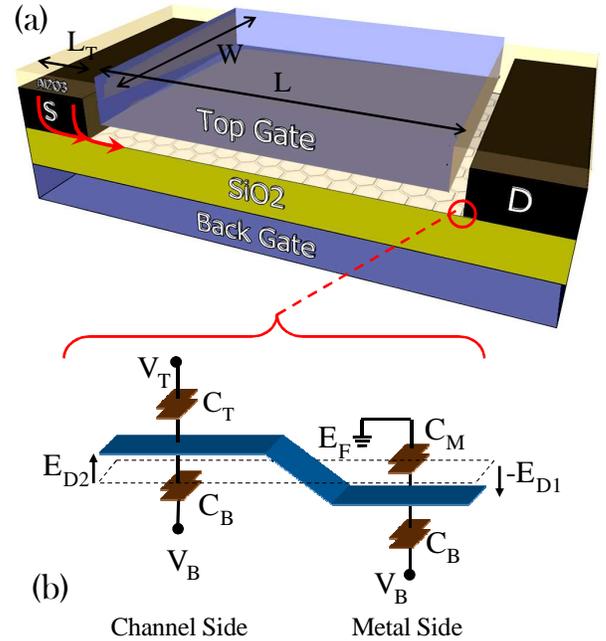}}
\caption{\footnotesize  {\bf(a)} Top/bottom gated graphene structure, and {\bf(b)} modeled potential profile across metal coated part and channel part of graphene sheet. With $C_{T}$/($C_{B}$) we denote the top/(bottom) gate capacitances, whereas $C_{M}$ accounts for the interfacial capacitance between metal and graphene. The positive/(negative) sign of the difference $E_{F}-E_{D}$ between the Fermi level $E_{F}$ and the Dirac surface $E_{D}$ accounts for n/(p) doping of the respective region.}
\label{fig1}
\end{figure}

Graphene is sensitive to external perturbations due to its all-surface and zero volume nature \cite{schedin07}. Charge transfer between metal and graphene due to a workfunction difference dopes the underlying graphene \cite{giovannetti01}. Contacts, therefore, introduce a built-in electrostatic junction within graphene, which was observed experimentally using scanning tunneling microscopy \cite{zhao01}. A distinct experimental signature was the asymmetry of resistance in back-gated devices \cite{huard01,cayssol01}. The sign of this asymmetry reflects the doping of the graphene underneath the metal. Recently, devices with top and bottom gating schemes, as shown in Fig.\,\ref{fig1}, were experimentally realized \cite{chen01,kim02}. Measurement of resistance vs. top gate voltage ($V_T$) revealed an asymmetry, whose sign and magnitude were modulated by the back gate voltage ($V_B$). This observation strongly suggests that the graphene doping underneath the metal was substantially modulated by the back gate voltage. 

It is instructive to recall that in standard metal-semiconductor junction theory, the Schottky barrier height is given by the difference between the metal workfunction, $\phi_{M}$, and the electron affinity, $\chi_{S}$, of the semiconductor, when there is no Fermi level pinning. However, the metal-graphene binding is comparatively weaker. The nature of the interfacial metal-graphene chemical bonding is still a subject of theoretical study \cite{vanin01}. Electrically, this interfacial layer can be modelled with an interfacial capacitance $C_{M}$. If $C_{M} \gg C_{B}$ and  $C_{M} \gg C_{qm}\approx e^{2}D(E_{F}-E_{D1}) $, where $D$ stands for the density-of-states, then one expects the Dirac point of graphene $E_{D1}$ to be stationary with respect to $\phi_{M}$. The above limit describes the standard metal-semiconductor junction, since $C_{M}$ is generally significantly larger than the semiconductor capacitance. If  $C_{M}\leq C_{qm}$, it is possible to modulate $E_{D1}$ with an applied voltage $V_{B}$. Consideration of this fact is key to explaining the experimental results \cite{chen01} of the device shown in Fig.\,\ref{fig1}.

\vspace{12pt}

\textbf{The model}: From Gauss's law, the electrostatic equations governing the graphene underneath the contact (region $1$) and top gate (region $2$) are
\begin{eqnarray}
C_{M}(E_{D1}+\delta\phi)+C_{B}(E_{D1}-eV_{B})=e^{2}n_{1}\\
C_{T}(E_{D2}-V_{T})+C_{B}(E_{D2}-eV_{B})=e^{2}n_{2},
\label{elec}
\end{eqnarray}
where $\delta\phi\equiv \phi_{M}-\phi_{G}$ is the workfunction difference between metal and graphene, and $en_{i}=E_{Di}^{2}/\pi(\hbar v_{f})^{2}$. Note that $E_{F}$ is taken to be zero as reference. The transition between region $1$ and $2$ is described by an analytical screening model \cite{zhang02}, assuming a linear graded junction. The junction resistance, ${\cal R}_{junc}$, is then computed quantum mechanically using a previously developed mode space non-equilibrium Green function method for graphene \cite{Low01}. For our ballistic study, the effect of temperature impacts our results only through the thermal smearing due to Fermi-Dirac distribution function, which is included in our study.

\begin{figure}[htps]
\centering
\scalebox{0.47}[0.47]{\includegraphics*[viewport=90 140 600 490]{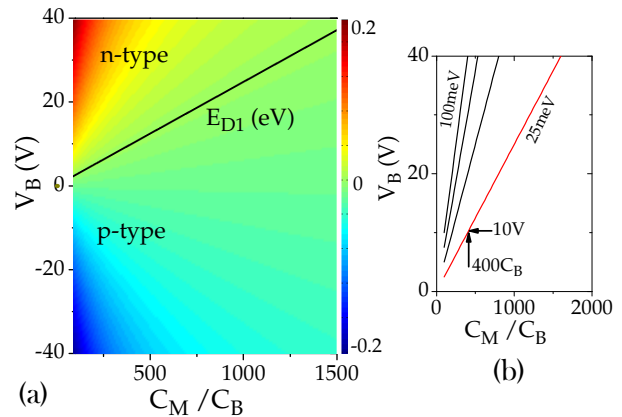}}
\caption{\footnotesize {\bf(a)} Surface potential (left), and {\bf(b)} modulation of the neutrality condition (right) of the metal coated side of graphene. The doping of the metal coated side of graphene is modulated with the backgate $V_{B}$ and depends also on the value of the interfacial capacitance $C_{M}$ and the work function difference $\delta\phi$. The crossing line(s) in both figures reflect(s) zero surface potential (Dirac lines). In the inset (right), the modulation of the neutrality condition of the metal coated side of graphene is applied, in direct correspondence to the equation $C_{M}\delta\phi-C_{B}V_{B}=0$}
\label{fig2}
\end{figure}

\begin{figure*}[htps]
\centering
\scalebox{0.55}[0.55]{\includegraphics*[viewport=60 40 700 580]{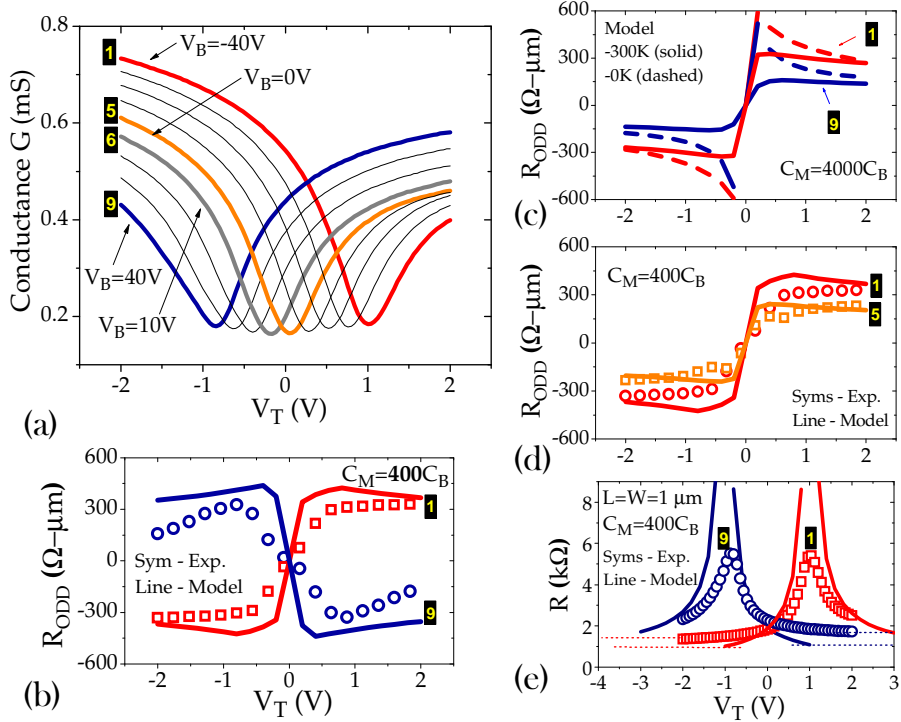}}
\caption{\footnotesize Study of the conductance asymmetry of the Chen/Appenzeller experiments \cite{chen01}. {\bf(a)} Experimental $G-V_{T}$ curves for different back gate voltages  $V_{B}$, ranging from $V_{B}=-40\,V$ to $V_{B}=40\,V$. {\bf(b)} Comparison of model prediction versus experiment for the odd part of resistance (${\cal R}_{odd}$) vs top gate voltage $V_{T}$ for $V_{B}=-40\,V$ and $V_{B}=40\,V$, assuming an interfacial capacitance $C_{M}=400C_{B}$, and a temperature of  $T=300\,K$. {\bf(c)} Model prediction of (${\cal R}_{odd}$) for the same voltages for an elevated interfacial capacitance $C_{M}=4000C_{B}$, for  $T=300\,K$ (solid line) and $T=0\,K$ (dashed line). {\bf(d)} Odd part of resistance (${\cal R}_{odd}$) vs $V_{T}$ for  $V_{B}=0\,V$ and $V_{B}=-40\,V$ with $C_{M}=400C_{B}$ at $T=300\,K$. {\bf(e)} Experimental $R-V_{T}$ curves for  $V_{B}=-40\,V$ and $V_{B}=40\,V$, versus our NEGF model, assuming an interfacial capacitance $C_{M}=400C_{B}$, a work function difference  $\delta\phi=0.025\,eV$ and a temperature of $T=300\,K$. }.  
\label{fig3}
\end{figure*}

In the experiment \cite{chen01}, the gate capacitances are known, i.e. $C_{B}\approx 1.15\times 10^{\mbox{-}4}\,Fm^{\mbox{-}2}$ and $C_{T}=42C_{B}$. On the other hand, $C_{M}$ is a quantity to be determined. Ti/Pd/Au is used for the metal contacts. For our calculations, we assumed $\delta\phi\approx 25\,meV$, a value that is sensible for our experimental metal stack \footnote{Literature on metal workfunction on graphene reports wide range of $\delta\phi$, suggesting that experimental condition and surface physics introduce variability. Experiments have provided indication about the metal-graphene function difference, based on quantitative $R_{ODD}$ analysis \cite{huard01}, and photocurrent study \cite{mueller01}.  From a computational standpoint, DFT studies \cite{giovannetti01,vanin01,giovannetti01} have provided support for a spectrum of work function differences in a range $\delta\phi=[-0.3,0.5]\,eV$ for several different metal compositions.}. Fig.\,\ref{fig2}a plots $E_{D1}$ as a function of $V_{B}$ and $C_{M}$. As expected, $V_{B}$ modulates the doping in region $1$, with greater ease when $C_{M}$ is smaller. In the experiments, it was observed that the conductance as function of $V_{T}$ exhibits the least asymmetry when $V_{B}\approx 10\,V$ (see Fig.\,\ref{fig3}a). This suggests that when $V_{B}\approx 10\,V$, $E_{D1}\approx 0$ (the charge neutrality point), which then also allows us to pin down $C_{M}$ to be $C_{M}\approx 400 C_{B}$. On the other hand, a different choice of $\delta\phi$ would correspondingly yield a different $C_{M}$, as illustrated in Fig.\,\ref{fig2}b.

\vspace{12pt}

\textbf{Results}: The measured conductance, $G(V_{T})$, for different values of $V_{B}$ is shown in Fig.\,\ref{fig3}a. The observed asymmetry in $G(V_{T})$ changes sign at about $V_{B}=10\,V$. To facilitate comparison between experiment and a ballistic theory, we extract the odd component of the resistance \cite{huard02} from experiment, given by ${\cal R}_{odd}\equiv \tfrac{1}{2}[{\cal R}(\delta V_{T})-{\cal R}(-\delta V_{T})]$, where $\delta V_{T}$ is $V_{T}$ with respect to the Dirac point voltage. The quantity ${\cal R}_{odd}$ then allows for quantitative comparison between the experimentally measured $G(V_{T})$ and the numerically calculated ${\cal R}_{junc}(V_{T})$. Although interface charge, moisture, and chemicals in the vicinity may impact the transport properties of the graphene devices, these effects can be minimized by careful control of the fabrication process and the measurement conditions so that any asymmetry can be attributed entirely to the graphene p-n junction.

In Fig.\,\ref{fig3}b, the striking experimental observation of asymmetry inversion, as it is observed in ${\cal R}_{odd}$ vs $V_{B}=-40\,V$ and $V_{B}=40\,V$, is compared with our modeled ${\cal R}_{odd}$ for a work function difference $\delta\phi=0.025\,eV$ and an interfacial capacitance $C_{M}=400C_{B}$. As elucidated previously, modulation of the doping of graphene underneath the metal is possible because $C_{M}$ is not large enough to completely dominate over $C_{qm}$. In fact, for a moderate carrier concentration of $1\times 10^{12}\,cm^{\mbox{-}2}$, one  obtains $C_{qm}\approx 0.75C_{M}$. If one assumes that the metal-graphene interfacial layer is an air gap, $C_{M}=400C_{B}$ would then translate to a physical thickness of only $2\,\AA$. This is in good agreement with recent density functional studies \cite{boukhvalov01}, with a predicted metal-graphene binding distance of $\approx 3.5\,\AA$. The possibility of sign inversion is, however, conditional. For example, if we use a value of $C_{M}=4000C_{M}$ instead of $C_{M}=400C_{B}$, sign inversion of ${\cal R}_{odd}$ would not be observed within the $V_{B}$ range of interest, as shown in Fig.\,\ref{fig3}c. In Fig.\,\ref{fig3}d, we show that the increasing odd resistance with increasing $\left|V_{B}\right|$, as it is observed in the experiment for $V_{B}=0,-40\,V$, can also be captured with our simulation, using $C_{M}=400C_{B}$. 

Up till now, we only considered ${\cal R}_{junc}$ and its contribution to the asymmetric part of the contact resistance. Previous studies \cite{chen02,perebeinos01} of the intrinsic transport properties of graphene on substrate allow one to make reasonable estimates of the channel resistance, ${\cal R}_{cha}$, by including contributions due to acoustic/optical phonons and substrate-induced remote phonons. However, their contributions relative to ${\cal R}_{junc}$ are not as significant \footnote{For example, acoustic phonons limited resistivity is only $30\Omega$ at room-temperature \cite{chen02}. Since our device has an aspect ratio of $W/L=1$. This contribution is not significant compared to ${\cal R}_{junc}$. }. A modest mobility of $\approx 500\, cm^{2}/Vs$ was extracted from the experiment in vicinity of the Dirac point, suggesting high levels of impurities \cite{adam07}. Here, we model the impurity limited resistivity with the Laudauer formula using a mean-free-path proportional in energy i.e. $\rho_{im}^{-1}=h/2e^{2}\times M^{-1}[L^{-1}+(\alpha E )^{-1}]$, where $\alpha$ is used to fit the mobility and $M$ is the number of modes normalized to $W$. In Fig.\,\ref{fig3}e, we compare the experimental resistance with the calculated sum ${\cal R}_{c}={\cal R}_{cha}+{\cal R}_{junc}$. By construction, our model does not capture the physics at the Dirac point. However, far away from the Dirac point, one observes an unaccounted for excess resistance in the experiment of $\approx 500\, \Omega$, the origin of which is the subject of subsequent discussion.

Other contributions to contact resistance include the current crowding effects due to access geometry and the presence of the metal-graphene interfacial layer. The latter implies that current has to tunnel across a dielectric layer, encapsulated in the electrical quantity $C_{M}=400C_{B}$. The tunneling resistivity $\rho_{tun}$ can be estimated using a model for quasi-bound electrons \cite{leggett01}, commonly used in the study of gate leakage current in semiconductor inversion layers \cite{hou01}. It is given by $\rho_{tun}=P\tau/e^{2}D$, where $P$ is the Wentzel Kramers Brillouin tunneling probability estimated to be $\approx 0.3$, and $\tau$ is known as the classical bounce time \cite{leggett01}, which in graphene is simply $\tau=2t_{g}/v_{f}$, with $t_{g}$ and $v_{f}$ being the thickness of graphene and Fermi velocity respectively. Here, we take $t_{g}$ to be the carbon-carbon bond length, $1.44\,\AA$. This yields us a tunneling resistance estimate of $\rho_{tun}\approx 5.2\times 10^{\mbox{-}6} \,\Omega$-$cm^{2}$ for a pair of contacts. This value is reasonably close to a recently reported \cite{nagashio02} experimental value of $\approx 5\times 10^{\mbox{-}6}\, \Omega$-$cm^{2}$. 

Expanding the prior analysis to the spatial distribution of carrier flow, the current crowding effects due to the metal-graphene acesss geometry lead to an effective electrical area for the contact where current flows, given simply by $W\times L_{T}$, where $L_{T}=\sqrt{\rho_{c}/\rho_{g}}$ is commonly known as the transfer length. With $\rho_{c}$ and $\rho_{g}$, we denote the specific contact resistivity and the sheet resistance of the graphene layer underneath, respectively. Using our estimated $\rho_{tun}\approx 5.2\times 10^{\mbox{-}6} \,\Omega$-$cm^{2}$ and a sheet resistance $\rho_{g}=1660\,\Omega/\square$ from the experiment \cite{chen01}, at a channel carrier concentration $n_{s}\approx 5\times 10^{\mbox{}12}\,cm^{\mbox{-}2}$, we find $L_{T}\approx 560\, nm$. Given the experiment to experiment variations in $\rho_{c}$ and $\rho_{g}$, this value is within a reasonable range \cite{nagashio02}.

\begin{figure}[t!]
\centering
\scalebox{0.30}[0.30]{\includegraphics*[viewport=0 0 750 600]{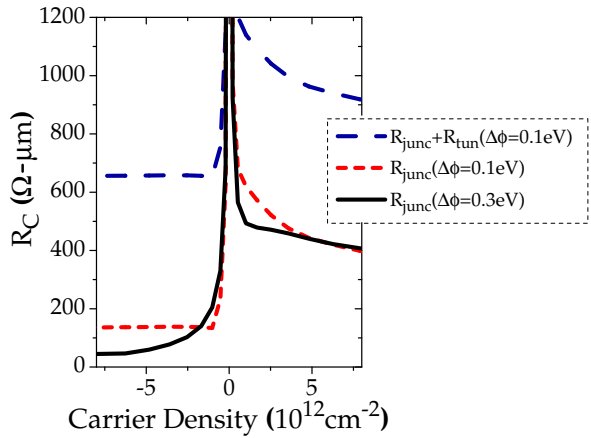}}
\caption{\footnotesize Estimates of various components of contact resistance in graphene transistors. With the long-dashed line, we denote our modeled ballistic prediction for the junction resistance plus the tunneling resistance ${\cal R}_{c}={\cal R}_{junc}+{\cal R}_{tun}$, for a pair of interfaces. After subtraction of ${\cal R}_{tun}$, we get the ballistic result for the lower bound of the contact resistance ${\cal R}_{junc}$ (short-dashed line), assuming a work function difference $\phi_{M}- \phi_{G} = 0.1\, eV$.  Last, an increase of the work function difference to $\phi_{M}- \phi_{G} = 0.3\, eV$ reduces further our ballistic predicition for the junction resistance ${\cal R}_{junc}$ (solid line). (The region near  $n_{s}=0$ should be disregarded since it is strongly affected by spatial potential fluctuations, which are not considered in our model.) }
\label{fig4}
\end{figure}

\vspace{12pt}

\textbf{Perspective}: Recently reported contact resistance values lie in the range ${\cal R}_{c}\approx600- 10^{\mbox{}4}\,\Omega$-$\mu m $ \cite{blake01,kim02,nagashio02,russo01,venugopal01,liu10} (see Suppl. Info.). These values are considerably above what is required for high performance transistors \cite{roadmap01}. In this section, we examine several issues related to the fundamental limit to ${\cal R}_{c}$.

In Fig.\,\ref{fig4}, we consider the contact resistance for a pair of interfaces ${\cal R}_{c}$, incorporating both the tunneling component ${\cal R}_{tun}= 520 \,\Omega$ that we extracted previously, and the junction component ${\cal R}_{junc}$ assuming a work function difference $\phi_{M}- \phi_{G} = 0.1\, eV$ (long-dashed line).  It is expected that ${\cal R}_{tun}$ can vary considerably from experiment to experiment, due to different contact materials, interface conditions, etc. \cite{blake01,kim02,nagashio02,russo01,venugopal01,liu10,xia01}. Significant reductions in ${\cal R}_{tun}$ will be crucial, and the contact metal and deposition and annealing conditions will be critical factors. Recently, it was demonstrated that a low power plasma $O_{2}$ treatment prior to metal deposition is beneficial in improving ${\cal R}_{tun}$ \cite{robinson01}. Another very recent study demostrates that ${\cal R}_{tun}$ is temperature dependent, possibly due to an expansion of graphene-metal distance  \cite{xia01}.

The ballistic junction component ${\cal R}_{junc}$, on the other hand, seems to have a more universal nature, since it is limited by the electrostatics condition, and the number of conducting channels bottleneck \cite{Low01}. To illustrate this, we plotted ${\cal R}_{junc}$ for $\phi_{M}- \phi_{G} = 0.1\, eV$ and $0.3\, eV$ in Fig.\,\ref{fig4}. For channel carrier concentration $n_{s}<0$, a lower resistance plateau forms. The resistance value of this plateau is described by the quantum contact resistance $R_{Q}=h/2e^{2}\times M^{-1}W$ \cite{Datta01}, where $M$ is the number of current-carrying modes, and this limit is imposed by $\Delta\phi$ at the metal side of the junction. Indeed for $\Delta\phi\,=\,0.1\,eV$ (short-dashed line), one finds $R_{Q}\approx 134 \,\Omega$-$\mu m$, while for $\Delta\phi\,=\,0.3\,eV$ (solid line) one finds $R_{Q}\approx 45\,\Omega$-$\mu m$. In the latter case, however, for moderate negative values of $n_{s}$, where the plateau hasn't been reached, quantum resistance $R_{Q}$ is limited by the number of modes in the channel, and $R_{Q}=h/2e^{2}\times1/2\sqrt{\pi/n_{s}}$ holds valid instead. The elevated right branch, when $n_{s}>0$, is due to the effect of interband tunneling. Selection of appropriate metal workfunction or approaches to engineer sharp $p$-$n$ junction, i.e. chemical doping \cite{avouris01}, are promising directions.

In conclusion, we have proposed a model that explains the gate dependent resistance asymmetry observed in experiments, and provides increased understanding of the different components of the contact resistance in graphene transistors. We show that the existence of an interfacial layer between the metal and graphene can explain the back gating of the contacts observed by Chen and Appenzeller \cite{chen01}. The importance of such an interfacial layer to devices was also pointed out in two very recent studies. Robinson et al. show that residual photoresist following lithography can lead to high contact resistance \cite{robinson01}, and Xia et al. explain the temperature dependence of the contact resistance in terms of the graphene to metal distance \cite{xia01}. Our work and these studies point to the need for further understanding of the metal-graphene interfacial layer as a prerequisite for engineering a good contact resistance for graphene electronics.

\emph{Acknowledgement:} This work has been supported by the Nanoelectronics Research Initiative through the INDEX center, and by the Focus Center for Material Structures and Devices. Computational support was provided by the Network for Computational Nanotechnology (NCN).

\end{document}